\documentclass[aps,prb,onecolumn,superscriptaddress,floatfix,showpacs,amsmath,amssymb,nofootinbib,longbibliography,nobibnotes,noeprint]{revtex4-2}

\usepackage{graphicx,color}
\usepackage{dcolumn}
\newcolumntype{.}{D{.}{.}{-1}}
\usepackage{bm}
\usepackage{epstopdf}
\usepackage{subfigure}
\usepackage{csquotes}
\usepackage[normalem]{ulem}
\usepackage{xcolor}
\usepackage{verbatim}
\usepackage{parskip,stackengine}
\usepackage{soul,xr}

\newcommand{\rk}[1]{\textcolor{black}{#1}}
\newcommand{\rev}[1]{\textcolor{black}{#1}}
\newcommand{\sau}[1]{\textcolor{black}{#1}}

\newcommand{\css}{$C_{\rm 66}$}
\newcommand{\ns}{$\chi$} 
\newcommand{\nst}{$\widetilde{\chi}$}

\newcommand{\chier}{$\chi^{er}$}
\newcommand{\chierz}{$\chi^{er}_{\rm 0}$}
\newcommand{\chiy}{$\chi^{sh}$}
\newcommand{\chiyc}{$\widetilde{\chi}^{sh}$}
\newcommand{\chiyb}{$\chi^{sh}$}

\newcommand{\tc}{$T_{\rm C}$}
\newcommand{\tn}{$T_{\rm N}$}
\newcommand{\ts}{$T_{\rm S}$}

\newcommand{\ter}{$T_{\rm 0}^{er}$}
\newcommand{\ty}{$T_{\rm 0}^{sh}$}
\newcommand{\ttil}{$\widetilde{T}^{sh}$}  

\newcommand{\la}{LaFeAsO}
\newcommand{\lao}{LaFeAsO$_{1-x}$F$_{x}$}
\newcommand{\ba}{BaFe$_2$As$_2$}
\newcommand{\baco}{Ba(Fe,Co)$_2$As$_2$}
\newcommand{\bacox}{Ba(Fe$_{1-x}$Co$_x$)$_2$As$_2$}
\newcommand{\etal}{\textit{et~al.}}

\begin{document}

\title{Nematicity in LaFeAsO single crystals studied by elastoresistance, high-resolution thermal expansion and shear-modulus measurements}

\author{X. C. Hong$^\ddagger$}\altaffiliation{Both auhors contributed equally.}
\affiliation{Leibniz-Institut f\"{u}r Festk\"orper u. Werkstoffforschung (IFW) Dresden, Helmholtzstr. 20, 01069 Dresden, Germany}
\affiliation{Fakultät für Mathematik und Naturwissenschaften,
Bergische Universität Wuppertal, Gaußstraße 20, 42119 Wuppertal, Germany}\email{hongxc@cqu.edu.cn}

\author{S.~Sauerland}\altaffiliation{Both auhors contributed equally.}
\affiliation{Kirchhoff Institute for Physics, Heidelberg University, D-69120 Heidelberg, Germany}

\author{L. Wang}
\affiliation{Kirchhoff Institute for Physics, Heidelberg University, D-69120 Heidelberg, Germany}

\author{F. Scaravaggi}
\affiliation{Leibniz-Institut f\"{u}r Festk\"orper u. Werkstoffforschung (IFW) Dresden, Helmholtzstr. 20, 01069 Dresden, Germany}

\author{A.~U.~B. Wolter}
\affiliation{Leibniz-Institut f\"{u}r Festk\"orper u. Werkstoffforschung (IFW) Dresden, Helmholtzstr. 20, 01069 Dresden, Germany}

\author{R.~Kappenberger}
\affiliation{Leibniz-Institut f\"{u}r Festk\"orper u. Werkstoffforschung (IFW) Dresden, Helmholtzstr. 20, 01069 Dresden, Germany}

\author{S.~Aswartham}
\affiliation{Leibniz-Institut f\"{u}r Festk\"orper u. Werkstoffforschung (IFW) Dresden, Helmholtzstr. 20, 01069 Dresden, Germany}

\author{S. Wurmehl}
\affiliation{Leibniz-Institut f\"{u}r Festk\"orper u. Werkstoffforschung (IFW) Dresden, Helmholtzstr. 20, 01069 Dresden, Germany}

\author{S. Sykora}
\affiliation{Leibniz-Institut f\"{u}r Festk\"orper u. Werkstoffforschung (IFW) Dresden, Helmholtzstr. 20, 01069 Dresden, Germany}

\author{F. Caglieris}
\affiliation{Leibniz-Institut f\"{u}r Festk\"orper u. Werkstoffforschung (IFW) Dresden, Helmholtzstr. 20, 01069 Dresden, Germany}

\author{B. B\"{u}chner}
\affiliation{Leibniz-Institut f\"{u}r Festk\"orper u. Werkstoffforschung (IFW) Dresden, Helmholtzstr. 20, 01069 Dresden, Germany}
\affiliation{Institut für Theoretische Physik and Würzburg-Dresden Cluster of Excellence ct.qmat,
Technische Universität Dresden, 01062 Dresden, Germany}

\author{C. Hess}
\affiliation{Leibniz-Institut f\"{u}r Festk\"orper u. Werkstoffforschung (IFW) Dresden, Helmholtzstr. 20, 01069 Dresden, Germany}
\affiliation{Fakultät für Mathematik und Naturwissenschaften,
Bergische Universität Wuppertal, Gaußstraße 20, 42119 Wuppertal, Germany}

\author{R. Klingeler}\email{klingeler@kip.uni-heidelberg.de}
\affiliation{Kirchhoff Institute for Physics, Heidelberg University, D-69120 Heidelberg, Germany}


\date{\today}

\begin{abstract}
Nematicity in LaFeAsO single crystals is studied by means of high-resolution thermal expansion, shear modulus, and elastoresistivity measurements.
A softening of the shear modulus $C_{\rm 66}$ towards the structural phase transition at $T_{\rm S}$ is observed. In addition, a similar Curie-Weiss-like divergence of the nematic susceptibilities is found in the temperature dependence of both $\chi^{sh}$ and $\chi^{er}$, which are deduced from the shear modulus (sh) and the elastoresistivity (er) studies, respectively. These observations provide evidence for an electronic origin of nematicity in LaFeAsO.
The characteristic energy of the coupling between the lattice and the electronic degrees of freedom is deduced to $\sim$30~K. The comparison to corresponding measurements on BaFe$_2$As$_2$ single crystals reveals a very similar temperature dependence of the shear modulus but yields contrasting results for $\chi^{er}$ : In BaFe$_2$As$_2$, $\chi^{er}$ diverges similarly as the uncoupled nematicity deduced from the shear modulus data as it is expected from the underlying Landau theory. In contrast, the Weiss temperatures of $\chi^{er}$ and $\chi^{sh}$ are significantly different in LaFeAsO. This difference is at odds with the commonly anticipated theories of resistivity anisotropy and electronic nematicity in iron pnictides.



\end{abstract}

\maketitle

\section{Introduction}

Unconventional superconductivity is accompanied by a complex interplay of competing or intertwined degrees of freedom. The interaction responsible for Cooper pairing is usually believed to manifest itself as ordered phases or pertinent fluctuations in the stoichiometric parent compounds~\cite{Chubukov2012,Hirschfeld2011}.
In this context, electronic nematicity has been identified as an intriguing property that may enhance the superconducting transition temperature \tc ~\cite{Fradkin2010,Lederer2015,Fernandes2012}.
While nematic fluctuations and/or nematic phases have been observed, e.g., in cuprate superconductors~\cite{Lawler2010,Keimer2015} and other novel strongly correlated systems, the probably most prominent example is represented by iron-based superconductors (\rev{FeSC})~\cite{Johnston,Stewart}. In most of their underdoped compounds, the tetragonal symmetry between the $x$- and $y$-directions in the Fe-plane is spontaneously broken when the system is cooled below its structural transition temperature \ts . The structural distortion is usually accompanied or followed by the evolution of long-range antiferromagnetic order \tn . In case of \tn\ $<$ \ts , rotational C$_4$-symmetry of the crystal lattice as well as the discrete Z$_2$-symmetry lifting the degeneracy of the Ising-like electronic ground state are broken in the intermediate phase while time-reversal symmetry is still preserved. \rk{This intermediate phase} is hence \rk{named} nematic~\cite{Fernandes2012}. In this phase, electronic anisotropy is found to be impressively large compared to a rather small orthorhombic distortion \rk{which suggests} that the structural-nematic transition does not arise from a simple elastic instability~\cite{fisherRev}.


\rk{It is however challenging to determine the driving instability of nematicity since symmetry breaking affects all correlated degrees of freedom so that every elastic, electronic, spin, and orbital property becomes anisotropic at the structural-nematic phase transition. 
Studying the associated susceptibilities directly in the high-symmetry phase~\cite{cano2010,fernandes2014_nature} allows to disentangle their specific contributions. For example,} the spontaneous elastic strain evolving at \ts\ as orthorhombic distortion $\delta$ acts as a conjugate field to a nematic order parameter $\psi$~\cite{chu2012,Fernandes2012} so that investigating the corresponding elastic and electronic response to an applied external stress $\sigma$ by measurements of the elastoresistivity and the elastic shear modulus have become powerful tools to access the nematic susceptibility
~\cite{chu2012,fernandes2010,bendingReview,Boehmer2022review}.
\begin{equation}
    \rev{\left(\frac{\partial \psi}{\partial \sigma}\right)^{-1} \propto ~ \widetilde{\chi}^{-1} = \chi^{-1} - C(\lambda)~}.
	\label{eq:nem_susc}
\end{equation}


In the literature~\cite{bendingReview, Kuo2016science}, \ns\ is often referred to as bare or uncoupled nematic susceptibility whereas \nst\ is termed renormalized or actual nematic susceptibility. Elastoresistivity measurements probe the normalized in-plane resistivity anisotropy $\eta = (\rho_b - \rho_a) / \rho_{\rm avg}$\footnote{$\rho_{a}$ and $\rho_{b}$ are the resistivities along the tetragonal in-plane directions of the macroscopic sample; $\rho_{\rm avg} = (\rho_a + \rho_b)/2$.}, being a measure of the nematic order parameter, in dependence of strain $\epsilon$. It thereby detects the purely electronic contribution to the nematic susceptibility, i.e.,  \ns~\cite{chu2012,kuo2013prb}. On the other hand, \rev{the softening of the elastic shear modulus is directly associated with both \nst\ and \ns, which differ by a term $C(\lambda)$ which depends on the electron-lattice coupling $\lambda$. This term, to be discussed later, accounts for the reduction in the energy of the spontaneous orthorhombic distortion by nematic fluctuations via the electron-lattice coupling~\cite{Fernandes2012}.}



For \ba\ and FeSe, the mentioned experiments have confirmed the shear modulus \css\ as the elastic soft mode of the tetragonal-to-orthorhombic transition~\cite{fernandes2010,yoshizawa2012,bending122,ultrasound11,bending11}, and have verified the transition's electronic origin~\cite{chu2012,hosoi, tanatar,watson,Baek2015,Fedorov2016,Baek2020}.
However, the picture is yet incomplete as experimental results on '1111'-compounds are  scarce. The relevance of such studies stems from the fact that, while FeSe does not show long-range magnetic order at ambient pressure and in undoped '122'-systems the structural-nematic and the magnetic phase transition coincide, in '1111'-compounds the transitions are split and give rise to another appearance of a nematic phase~\cite{delaCruz2008,Luetkens2009,Hess2016}, \rev{presenting a unique testbed for intertwined orders}.
In addation, testing theoretical models of nematicity in this class of materials is also particularly relevant as the '1111'-family still holds the highest \tc\ among bulk \rev{FeSC} at ambient pressure~\cite{Wang2008high,Wu2009high,Cheng2009high}. Previous experimental studies on \la\ polycrystals showed clear indications of fluctuations of all relevant degrees of freedom well above \ts~\cite{Wang2009,Grafe2009,Maeter2009,Klingeler2010}. However, these investigations have so far been limited by constraints in single crystal growth. The recent developments in growing large high-quality single crystals~\cite{kappenberger2018} now enable detailed studies of the nematic properties in the '1111'-family of pnictides. Recent studies on single crystals show that nematic fluctuations promote spin fluctuations in magnetically ordered LaFeAsO~\cite{Ok2018} and foster the superconductivity in doped systems~\cite{Hong2020,Wuttke2022}. In this paper, we present a combined study of elastoresistivity, high-resolution thermal expansion and shear modulus measurements in order to gain more insight into the microscopic origin of nematicity in \la .

\section{Experimental}

Well-faceted \la\ single crystals grown by the solid state single crystal growth (SSCG) method~\cite{kappenberger2018} of 
of about (1.0 $\times$ 0.8 $\times$ 0.3) mm$^3$ have been investigated. Thermal expansion studies were performed by means of a three-terminal high-resolution capacitance dilatometer (Kuechler Innovative Measurement Technology) in a home-built set-up~\cite{kuechler2012, werner2017}. Temperature control was ensured by a Variable Temperature Insert of an Oxford magnet system~\cite{Spachmann2021}. The dilatometer's built-in repulsion force of two leaf springs allows to detwin the sample for an orientation along the [110]$_{t}$ direction~\cite{Boehmer_detwinning,kuechler2012}. The Young's modulus along this particular direction has been obtained by a three-point-bending (3PB) technique similar as presented in Ref.~[\onlinecite{bending122}]. In the dilatometer, a plate-like sample of about (1.5 $\times$ 1.5 $\times$ 0.2) mm$^3$ is positioned between three rods as it is shown in the inset of Fig.~\ref{TEbending}(c). Thus, the force of the springs acting along the [001]-direction bends the sample and its deflection is measured by the distance change of the dilatometer's gap. This enables calculating Young's modulus $Y_{\rm [110]}$ which is directly related to the elastic constant \css ~\cite{kityk1996}.


Single crystals from the same batch were used for elastoresistance measurements. Strain was generated by firmly gluing the prepared samples on the side of a piezoelectric actuator (PZT)~\cite{chu2012}. The crystals were cut to rectangular shape with the long sides along the [110]$_{t}$ or [100]$_{t}$ directions. In order to achieve efficient strain transmission, the samples were cleaved to a thickness of around 20~$\mu$m. Electrical contacts were made directly on the fresh surfaces with silver paint. Resistance was measured by a standard four-point $dc$ technique.

\section{Results}

\begin{figure}
	\includegraphics[width=0.7\columnwidth,clip] {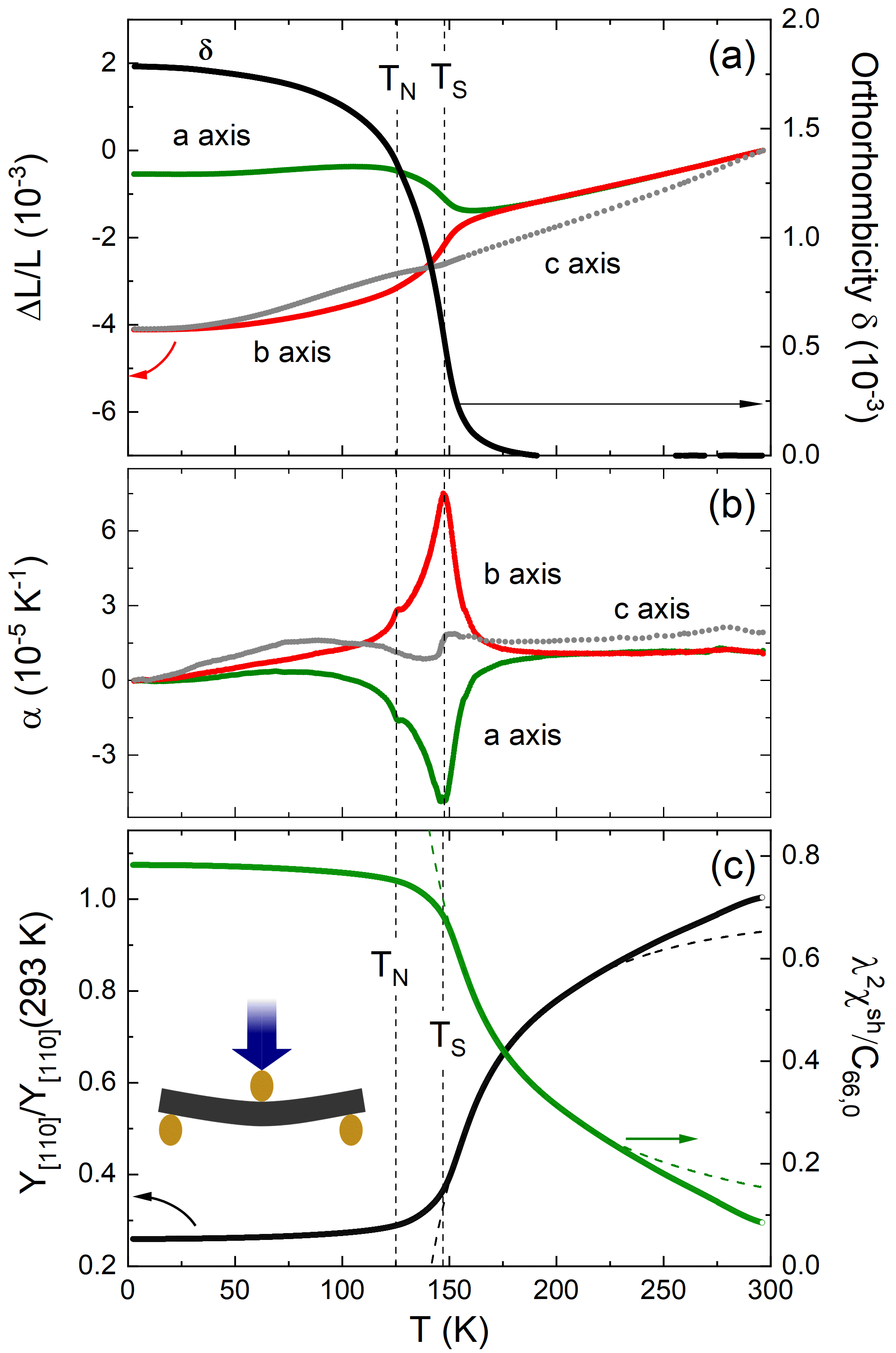}
	\caption{(a) Relative length changes $\Delta L/L$ of the \rk{three} main crystallographic axes (left scale) and orthorhombic distortion $\delta = (a-b)/(a+b)$ (right scale) \textit{vs.} temperature. Structural and magnetic phase transition temperatures, \ts\ and \tn , of this crystal are marked by dashed lines. (b) Corresponding uniaxial thermal expansion coefficients. (c) Young's modulus $Y_{\rm [110]}$ along the [110]$_{t}$ direction (left scale) and bare nematic susceptibility $\lambda^2 \chi^{sh} /C_{\rm 66,0}$ (right scale) together with Curie-Weiss fittings (dashed lines) according to Eq.~\ref{CW-law_Y}. The inset shows a schematic sketch of the utilized 3PB-setup where the sample bends under an applied force (blue arrow).}
	\label{TEbending}
\end{figure}

The thermal expansion data shown in Fig.~\ref{TEbending}(a) illustrate the temperature dependence of the lattice parameters, where $b$ denotes the shorter in-plane direction of the detwinned crystal. The relative length changes along the $a$ axis have been calculated from the difference of length changes along the detwinned and the twinned [100]$_{t}$ direction \rev{(see Fig.~A2 in the Appendix)}. While in the high-temperature tetragonal phase upon cooling, the sample shrinks along all three axes, the orthorhombic splitting of $a$ and $b$ signals the structural transition at \ts\ = 148~K. As shown in Fig.~\ref{TEbending}(a), an orthorhombic distortion $\delta = (a-b)/(a+b)$ evolves rather smoothly with a large \rk{precursor regime}. The orthorhombic distortion $\delta$ represents the order parameter of the tetragonal-to-orthorhombic phase transition. Note that the evolution of orthorhombicity is affected by finite pressure applied in the experimental setup but the general behavior agrees well to neutron data~\cite{Qureshi} on polycrystalline \lao~\cite{Wang2019,Scaravaggi2021}.

\rk{The temperature dependence of} Young's modulus $Y_{\rm [110]}$, which above \ts\ probes the elastic shear modulus \css , is depicted on the left ordinate in Fig.~\ref{TEbending}(c). Fairly similar to what has been observed in BaFe$_{\rm 2}$As$_{\rm 2}$~\cite{bending122} and FeSe~\cite{bending11} (see also Fig.~\ref{Y_scaled} in the Appendix), \rk{$Y_{\rm [110]}$ shows a strong decrease towards \ts , thereby reflecting a clear softening of \css . This implies that} \css\ is the soft mode of the orthorhombic distortion also in \la . At \ts , the curve flattens in a rather continuous manner and the inflection point at 154.5~K marks a lower limit to observe the softening. \rk{Note, that inhomogeneous stress applied to the sample via the 3PB setup (see Fig.~\ref{TEbending}) may smear out the anomaly due to strong strain dependencies of \ts . }
Specifically, measurements under varying uniaxial pressure along the [110]$_{t}$ direction show a shift of \ts\ and, in \ba , breaking of the C$_4$-symmetry already above \ts~\cite{blomberg,dhital,hu,wangPRB2018,Scaravaggi2021}. 
\rk{We also note that, $Y_{\rm [110]}$(\ts) remains finite at \ts . This observation contrasts theory predictions of vanishing \css\ but is typically observed in 3PB and in ultrasound experiments on \ba ~\cite{bending122,fernandes2010} and FeSe~\cite{bending11, ultrasound11}.} 
Below \ts , $Y_{\rm [110]}$ reaches a constant value which may be ascribed to the motion of structural domain walls in the orthorhombic phase~\cite{bendingReview, schranz}.

As shown in Fig.~\ref{TEbending}(a), the orthorhombic distortion parameter $\delta$ evolves continuously as similarly observed for \la\ polycrystals~\cite{Qureshi,Wang2009}. In addition, the temperature dependence of $Y_{\rm [110]}$ and accordingly of \css\ above \ts\ [see Fig.~\ref{TEbending}(c)] indicates that the structural transition is not barely driven by an elastic instability. In a proper ferroelastic scenario the shear modulus is supposed to decrease linearly with temperature~\cite{cano2010} which is not observed in the data at hand.

\begin{figure}
\includegraphics[width=1.0\columnwidth,clip]{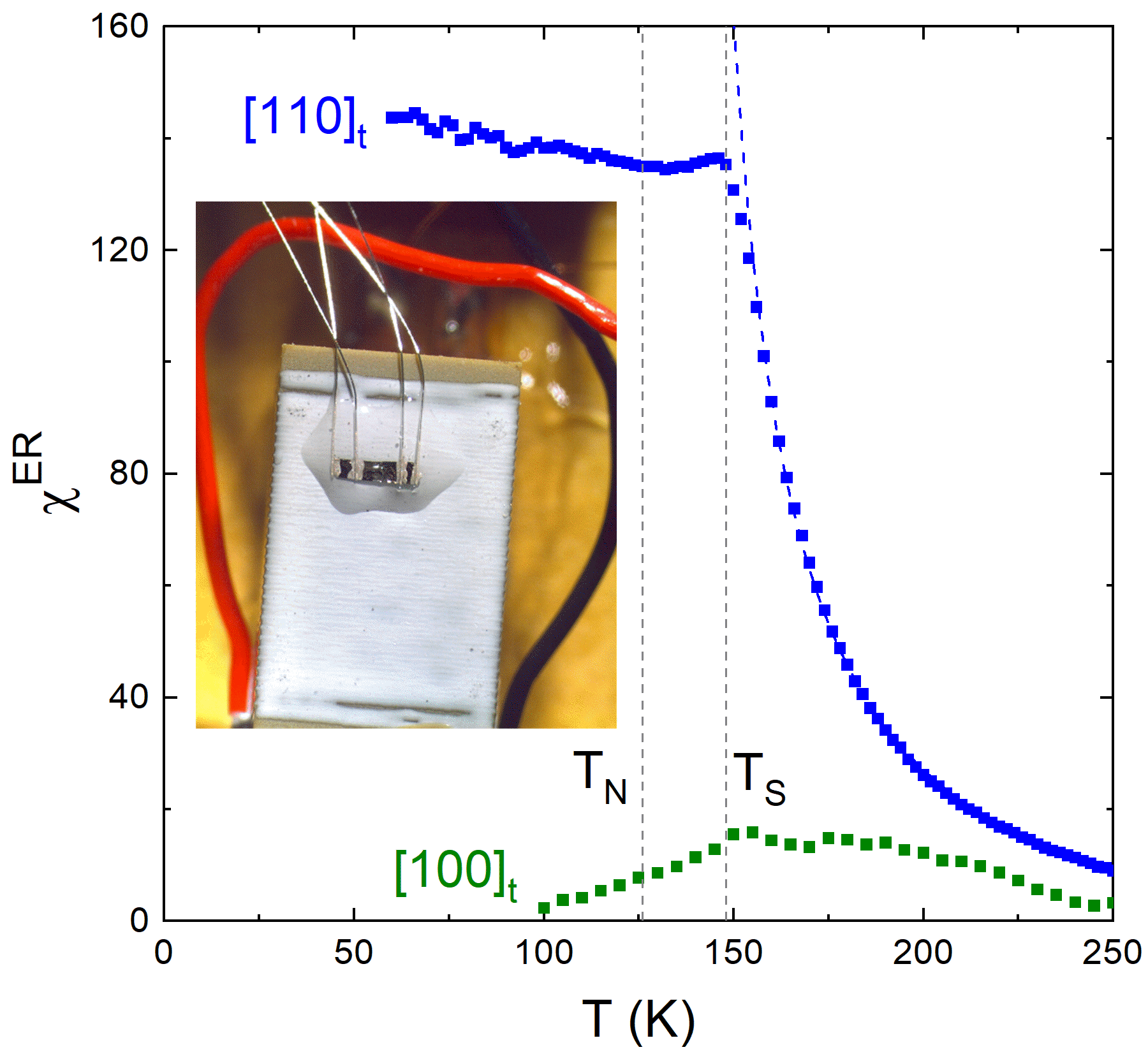} 
\caption{Temperature dependence of the nematic susceptibility measured by elastroresistance $\chi^{er} = - d\eta / d\epsilon$. $\chi^{er}$ in the $B_{\rm 2g}$- and the $B_{\rm 1g}$-symmetry channels was investigated by applying strain $\epsilon$ along the [110]$_{t}$ and [100]$_{t}$ crystal directions, respectively. Dashed lines indicate \ts\ and \tn . The blue dashed line shows Curie-Weiss-fitting according to Eq.~\ref{CW-law_er}. The inset shows a photograph of a crystal glued on the surface of a PZT stack.}\label{ER}
\end{figure}

The in-plane resistivity anisotropy $\eta$ 
probes the nematic order parameter in dependence of elastic strain $\epsilon$~\cite{chu2012}
\begin{equation}
	\chi^{er} = - d\eta / d\epsilon \propto - d\psi / d\epsilon.
\end{equation}
Depending on the relative direction of the sample axis with respect to the strain direction, i.e., $\epsilon$ either $\|$ [110]$_{t}$ or $\|$  [100]$_{t}$, respectively, $\chi^{er}$ is related to the $B_{\rm 2g}$- or the $B_{\rm 1g}$-symmetry channel~\cite{kuo2013prb}. The experimental data shown in Fig.~\ref{ER} indeed show strong differences for both channels. \rk{For $\epsilon \| [110]_{t}$, $\chi^{er}$ is large and diverges towards a kink at \ts\ while it is damped and featureless for $\epsilon \| [100]_{t}$.} We conclude that the nematic order parameter only develops in the diagonal $B_{\rm 2g}$-channel which agrees with the shear modulus results as \css\ refers to the same symmetry. \rk{This finding matches with those} for other \rev{FeSC}~\cite{Kuo2016science}. 
Note, the sign of $\chi^{er}$ is positive for LaFeAsO like in Ba(Fe$_{1-x}$Co$_x$)$_2$As$_2$ but \rk{$\chi^{er}$ is negative} in FeSe~\cite{Kuo2016science,hosoi,tanatar}. 
\rk{The following discussion will focus on the data above \ts\ since the elastroresistance is known to be dominated by domain effects below \ts~\cite{fisherRev}.}


Similar to other \rev{FeSC}~\cite{chu2012,bendingReview}, the structural (and nematic) transition is described by means of a pseudo-proper ferroelastic approach where the free energy is given by
\begin{equation}
	F = \frac{\chi^{-1}}{2}\psi^2 + \frac{C_{\rm 0}}{2}\epsilon^2 - \lambda\psi\epsilon - \sigma \epsilon .
	\label{free_energy}
\end{equation}
In this Landau expansion, the electronic order parameter $\psi$ is bilinearly coupled via $\lambda$ to the elastic strain $\epsilon$ while $\sigma$ refers to an externally applied stress. $\chi^{-1}$ stands for the inverse bare nematic susceptibility in absence of coupling while $C_{\rm 0}$ ($=C_{\rm 66,0}$) is the inverse bare elastic susceptibility that accounts for the elastic energy by $C_{\rm 0} \epsilon^2 / 2$ in the absence of coupling. Minimizing the free energy with respect to $\psi$ and $\epsilon$ reveals that the response of the nematic order parameter $\psi$ to elastic strain $\epsilon$ is a direct measure of the \sau{uncoupled} nematic susceptibility~\cite{chu2012}
\begin{equation}
	\left. \frac{d\psi}{d\epsilon} \right|_{\epsilon = 0}  = \lambda\chi.
	\label{dpsi_deps}
\end{equation}
Hence, the observed divergence of $\chi^{er}$ \rev{provides evidence} for an electronic origin of nematicity in \la.

Moreover, minimizing $F$ yields that the shear modulus \css\ is renormalized by the electron-lattice coupling with $\widetilde{\chi}$ being the renormalized susceptibility~\cite{Fernandes2012,bendingReview}:

\begin{equation}
	C_{\rm 66} = \left( \frac{d\epsilon}{d\sigma} \right)^{-1} = C_{\rm 0} \left( 1 + \frac{\lambda^2 \widetilde{\chi}}{C_{\rm 0}} \right)^{-1} = C_{\rm 0} - \lambda^2\chi.
	\label{eq:c_eff}
\end{equation}

\begin{figure*}
	\centering
	\includegraphics[width=\textwidth,clip]{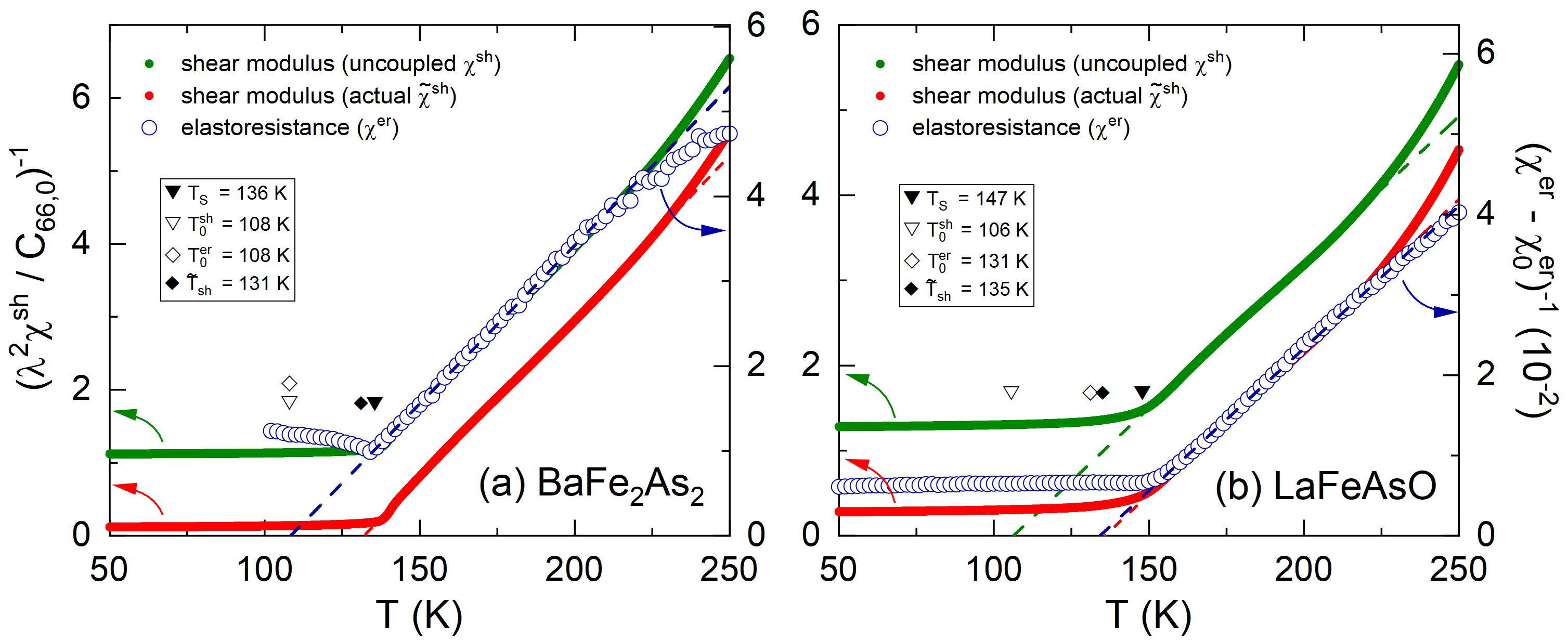}
	\caption{Temperature dependence of the inverse purely electronic nematic susceptibilities $(\lambda^2 $\chiyb$ /C_{\rm 66,0})^{-1}$ (left scale, green markers) and $($\chier$ - $\chierz$)^{-1}$ (right scale, blue markers) of (a) \ba\ and (b) \la\ obtained by shear modulus and elastoresistivity measurements, respectively. \chierz\ accounts for the intrinsic \rk{elasto}resistivity. In addition, $(\lambda^2 $\chiyc$ /C_{\rm 66,0})^{-1}$ (left scale, red markers) with the actual (renormalized) nematic susceptibility \chiyc\ is shown. The dashed lines represent Curie-Weiss fits to the data. Characteristic temperatures are indicated by grey markers (see the text).}
	\label{inv_susc}
\end{figure*}

In case of finite electronic-elastic coupling $\lambda$, the structural phase transition hence appears at the temperature where the experimentally obtained inverse elastic susceptibility \css\ as well as the inverse of the actual nematic susceptibility \nst\ vanish~\cite{Fernandes2012,bendingReview}. At this temperature, $\delta$ and $\psi$ become finite although the nematic instability itself appears at lower temperature $T_{\rm 0}$ at which \ns\ is expected to diverge~\cite{chu2012}. Eq.~\ref{eq:c_eff} also demonstrates that the experimentally determined shear modulus \css\ is directly linked to the renormalized nematic susceptibility \nst\ and also the bare electronic contribution \ns\ can be deduced~\cite{bendingReview}.

Normalized values $Y/Y(293~K)$ obtained by the 3PB-method [see Fig.~\ref{TEbending}(c), left scale] enable one to determine the uncoupled nematic susceptibility \chiyb\ in units of $\lambda^2/C_{\rm 66,0}$ using Eq.~\ref{eq:c_eff} and following Ref.~[\onlinecite{bendingReview}], i.e.,
\begin{equation}
	\lambda^2\chi^{sh}/C_{\rm 66,0} = 1 - C_{\rm 66} / C_{\rm 66,0} \approx 1 - Y_{\rm [110]} / Y_{\rm 0}.
	\label{eq:suscept_shear}
\end{equation}
\rk{Our treatment of the non-critical background $C_{\rm 66,0}$ and accordingly of $Y_{\rm 0}$ is based on the model for \ba~\cite{bending122} using the formula by Y.~P.~Varshni~\cite{Varshni1970} (denoted by $F_{\rm Varshni}$ in Eq.~\ref{CW-law_Y}). The procedure involves the determination of a factor $p^{sh}$ which links absolute values of the background and the present measurement and was also successfully applied for FeSe~\cite{bending11}.} 
With the assumption $\chi^{-1} = a (T-T_{\rm 0})$, usually utilized in Landau theories, Young's modulus data [Fig.~\ref{TEbending}(c), left scale] is well fitted by
\begin{equation}
	\frac{Y_{\rm [110]}}{Y_{\rm [110]}(293~K)} = p^{sh} F_{\rm Varshni} \left( 1 - \frac{\lambda^2 / a C_{\rm 66,0}}{T - T_0^{\rm sh}} \right) .
	\label{CW-law_Y}
\end{equation}
The resulting uncoupled nematic susceptibility \chiyb\ is visualized in Fig.~\ref{TEbending}(c) (right scale). 
Similarly, the susceptibility \chier\ from elastoresistance is fitted by 
\begin{equation}
	\chi^{er}  = \frac{p^{er} \lambda / a}{T - T_{\rm 0}^{\rm er}} + \chi^{er}_0
	\label{CW-law_er}
\end{equation}
where \chierz\ accounts for a background due to an intrinsic piezoresistivity and $p^{er}$ for the unknown proportionality factor linking the strain dependence of the nematic order parameter $\psi$ and the resistivity anisotropy $\eta$. Note the similar form of Eq. \ref{CW-law_Y} and \ref{CW-law_er} \rk{which both contain} three unknown fitting parameters.


In order to compare the critical behavior, the temperature dependencies of the inverse susceptibilities obtained from elastoresistivity and shear modulus measurements are shown in Fig.~\ref{inv_susc}(b), indicating a linear, i.e., Curie-Weiss-like decrease for both techniques in \la. In Fig.~\ref{inv_susc}(a), we additionally present the same quantities obtained in our experimental setups on \ba\ single crystals~\cite{SaiBa122,Ni2009}. \rk{These data} agree well with the previous literature~\cite{chu2012,Kuo2016science,yoshizawa2012,bending122}. As expected from Landau theory, for \ba\ we observe a linear, i.e., Curie-Weiss-like decrease upon cooling towards the nematic instability temperature $T_{\rm 0}$. A fit to the data yields \ty $= 108(6)$~K and \ter $=108(5)$~K for the shear modulus and the elastoresistivity data, respectively. \rk{We also show the} actual, i.e., renormalized, nematic susceptibility \chiyc\ which is obtained from shear modulus data by Eq.~\ref{eq:c_eff}. Obeying a Curie-Weiss-like behavior, \chiyc\ diverges at \ttil $=131(3)$~K at which $Y_{[110]}$ extrapolates to zero according to Eq.~\ref{CW-law_Y}.
These results \rk{shown in Fig.~\ref{inv_susc} demonstrate a match} of the bare mean field nematic critical temperatures obtained by two experimental techniques. \rk{Such a match was predicted by Landau theory} but has not been explicitly reported in the literature so far~\cite{Kuo2016science,bendingReview}. 
The behavior of \chier , \chiyb , and \chiyc\ is further illustrated by scaling $($\chier$-$\chierz$)^{-1}$ to $(\lambda^2 $\chiy$ /C_{66,0})^{-1}$ by a single proportionality factor as it has been done by proper adjustment of the ordinates in Fig.~\ref{inv_susc}. The scaling factor of about 0.009 may be attributed to the quantity $\lambda / p^{\rm er} C_{\rm 66,0}$. Using literature data on \chier~\cite{Kuo2016science} yields a very similar value.


For \la, fitting \chiyb\ and \chiyc\ yield the Weiss temperatures \ty $=106(9)$~K and \ttil $=135(3)$~K, respectively. \rk{These values are similar to those in \ba. Even the respective scaling factor $p^{\rm sh}$ is  almost the same.} In particular, normalizing the temperature axis to \ts\ of each material yields an almost identical decrease of $Y_{\rm [110]}/Y_{\rm [110]}(293~K)$ for both compounds. So the main difference between \ba\ and \la\ lies in the discrepancy of the mean field (\ttil) and actual (\ts) structural transition temperature which is larger in \la\ than in \ba\ while the cause of this discrepancy, at all, is unknown~\cite{bendingReview}.
From the difference \ttil $-$ \ty, we conclude that the characteristic energy scale of the coupling $\lambda^2/aC_{\rm 66,0}$ 
amounts to $\sim$30~K~\cite{Boehmer2022review}. This value is in accordance with the corresponding value for \ba\ (see above) as well as the literature values for FeSe~\cite{bending11} and \ba ~\cite{bending122} and somewhat smaller than the one for \ba\ reported in Ref.~[\onlinecite{yoshizawa2012}].
However, for the elastoresistance, the Curie-Weiss-analysis yields \ter $=133(7)$~K which is significantly higher than \ty . \rk{This difference cannot be explained by rationalization of experimental uncertainties since special care} has been taken to account for the influence of the selected fitting temperature range and systematic correlations between fitting parameters due to the unknown backgrounds. \rk{Our analysis} indicates, also for \ba, that fitting regimes closer to \ts\ tend to yield higher values for \ty\ and \ttil\ but smaller ones for \ter. However, as visible by the uncertainties, this effect does not explain the large discrepancy for \la.

In fact, \ter\ seems to match rather \ttil\ giving rise to scaling of $($\chier$-$\chierz$)^{-1}$ and $(\lambda^2$\chiyc$ /C_{\rm 66,0})^{-1}$ by a factor of about 0.011 as shown in Fig.~\ref{inv_susc}(c). So despite the differences of the Weiss temperatures, the \rk{corresponding} Curie constants $p^{\rm er} \lambda / a$ from elastoresistance as well as the ratio $\lambda / p^{\rm er} C_{\rm 66,0}$ from the scaling are very similar for \ba\ and \la. 

\section{Discussion}

\rk{Despite the strongly different} appearance of a long-range magnetically ordered phase, the evolution of Young's modulus $Y_{\rm [110]}$ and of the nematic susceptibility $\lambda^2$\chiy$/C_{\rm 66,0}$ [Fig.~\ref{TEbending}(c)] as well as of \chier\ in the $B_{\rm 2g}$-channel (Fig.~\ref{ER}) imply clear similarities in \la\ as compared to other \rev{FeSC} such as \ba~\cite{bending122, chu2012} and FeSe~\cite{bending11, hosoi}. \rk{Apparently, the absence of} static nematic order in \ba\ seems to have no significant impact on the softening of the shear modulus. A similar conclusion was drawn for FeSe where \chiy\ almost resembles the behaviour of 3\%Co-doped \ba\ both exhibiting \ts\ $\sim 90$~K whereas FeSe does not show long-range magnetic order at finite temperatures at all~\cite{bending11}.


Apart from these similarities, \rk{there is a clear discrepancy between the findings in LaFeAsO and \ba . In LaFeAsO, \chier\ may be scaled to \chiyc\ rather than to \chiyb , i.e., the Weiss temperature \ter $=133(7)$~K clearly exceeds \ty\ but rather matches \ttil\ (cf. Fig.~\ref{inv_susc}). This observation clearly contrasts the behaviour in \ba\ as shown in Fig.~\ref{inv_susc}(a) and the literature~\cite{bending122,Kuo2016science}.}



Our results on \la\ may be also compared to Co-doped \bacox\ which, for $x\neq 0$, exhibits a static nematic phase, i.e., \ts $>$\tn . While for $x=0$ and 0.025, Kuo \etal~\cite{Kuo2016science} determined \ter\ from elastoresistivity to be well below \ttil\ observed by shear modulus measurements~\cite{bending122}, \ter\ appears to be even larger than \ttil\ for higher doping (x=0.047).
As a potential reason, an overestimation of \ter\ in fitting due to low temperature deviations from Curie-Weiss behavior observed in these compounds has been suggested~\cite{bendingReview}. However, our data of both techniques and for both compounds emphasize a very clear Curie-Weiss behavior even down to temperatures close to \ts\ giving no indication for physics beyond a mean-field approach.

Moreover, Curie constants from elastoresistivity are nearly the same for \ba\ and \la\ so that the data do not show any significant enhancement of \chier\ for one of the compounds. Such an enhancement of the elastoresistivity 
has been observed in \baco\ upon approaching optimal doping and has been associated with renormalization of the quasiparticle effective mass by nematic quantum critical fluctuations~\cite{Kuo2016science}. However, we cannot exclude that the proximity of a quantum critical point causes the unexpectedly high \ter . In this case, our data would suggest an enhancement of the elastoresistance in \la\ already far below the superconducting regime placing \la\ closer to quantum criticality than \ba .
Apart from that, it is not fully understood how magnetic and orbital fluctuations microscopically influence the elastoresistivity and shear modulus. \rk{Such effects might differ for transport and thermodynamic properties.} 
In this respect, \chier\ is thought to be influenced by the detailed electronic structure and/or scattering processes~\cite{Kuo2016science,hosoi,tanatar}.

Therefore, especially the effect of disorder as a possible explanation for the resistivity anisotropy has already been widely addressed in the literature, both experimentally and theoretically~\cite{fernandes2011,blomberg,Breitkreiz2014,gastiasoro2014_dimers,gastiasoro2014_resisanis, allan2013_STM, ishida2013}. In particular, Kuo and Fisher argued that for undoped as well as for Co- and Ni-underdoped \ba , the evolution of the nematic susceptibility and the resulting Weiss temperatures \ter\ are essentially independent of disorder~\cite{kuo2014disorder}. Moreover, the magnitude of the elastoresistance coefficient 
is found to be similar for all optimally doped \ba\ compounds, \rk{suggesting that disorder emerging from different types of dopants is not at play~\cite{Kuo2016science}.} 
Related to this, our data for \ba\ and \la\ indicate no significant difference between their respective Curie constants $p^{\rm er} \lambda / a$ from elastoresistivity.
On the other hand, Gastiasoro~\etal~\cite{gastiasoro2014_resisanis} suggested that assuming a significant amount of disorder in the samples, anisotropic spin fluctuations provoked by orthorhombic symmetry breaking are locally pinned to impurities and, \rk{thereby,} establish defect states called nematogens. By this, the defects exhibit anisotropic scattering potentials which hence manifest themselves in an anisotropic resistivity.
Calculations~\cite{gastiasoro_diss} based on this model may even reproduce a Curie-Weiss-like divergence of d$\psi$/d$\delta$ ($\delta$: orbital order parameter accounting for orthorhombicity of the band dispersion) towards the bare magnetic phase transition temperature $T_{\rm N,0}$. In Ref.~[\onlinecite{gastiasoro2014_resisanis}], the authors are in particular pointing out that Kuo's investigations~\cite{kuo2014disorder} would rather refer to out-of-plane disorder and within the mentioned scenario, the anisotropy is mostly caused by spin fluctuations so that different kinds of impurities should be reflected rather in the average resistivity than its anisotropy.
Within the scenario of disorder, our findings would propose a larger effect of disorder in \la\ than in \ba\ which is consistent with expectations for systems that show split second-order nematic and magnetic transitions as in \la\ and underdoped \baco\ compared to simultaneous (or close-by) transitions as in \ba~\cite{fernandes2014_nature}. \rk{Recent} experimental investigations on optimally doped LaFeAsO$_{1-x}$F$_x$ polycrystals emphasize the high sensitivity of these compounds on magnetic impurities where already 0.5\% Mn/Fe substitution re-establishes a magnetically ordered, orthorhombic state similar as it appears in undoped \la~\cite{hammerath2014,Moroni2016,Moroni2017}. \rk{Strong electronic correlations have been suggested} to provoke enhanced RKKY-coupling between Mn defects and thereby localization of electrons~\cite{gastiasoro2016,Moroni2017}.

From a different perspective, also stoichiometry of the samples may affect the transport properties. While Ba122-systems are rather stoichiometric, 1111-systems are known to possibly exhibit oxygen deficiencies~\cite{wu2008_1111} which are discussed for single crystals of the present work~\cite{kappenberger2018} as a possible explanation for lower values of \ts\ and \tn\ compared to the respective values of polycrystals.

However, more experimental research as well as theoretical calculations are needed to clarify the importance and the role of disorder. Our study, therefore, indicates a good starting point for such a study opening up that the direct comparison of shear modulus and elastoresistivity on the doping evolution of \la\ single crystals might shed more light on these open questions.

\section{Summary}

\rk{We report measurements of the thermal expansion, the shear modulus, and the elastoresistivity in order to investigate the nematic phase in LaFeAsO single crystals.} Our data imply a clear softening of the shear modulus \css\ towards the structural phase transition at \ts\ and a similar Curie-Weiss-like divergence of the nematic susceptibility obtained with both techniques, i.e., shear modulus and elastoresistivity studies. We therefore conclude an electronic origin of nematicity in \la. A characteristic energy scale of $\sim 30$~K for the electron-lattice coupling $\lambda^2$/a$C_0$ is obtained.
The softening of \css\ does not show any significant difference between LaFeAsO and the iconic \ba\ when accounting for their different \ts , albeit the emergence of their corresponding nematic phase is significantly different. 
\rk{The Curie-Weiss-like divergence of the purely electronic nematic susceptibility probed by elastoresistivity in \ba\ and LaFeAsO, however, indicates a significant difference in both materials. In \ba , the mean-field divergence of \ns\ occurs at a temperature consistent with the corresponding temperature determined from shear modulus data. While this is expected from the Landau theory, in LaFeAsO this divergence in \ns\ appears at a significantly higher temperature than in the shear modulus data.} Specifically, the critical part of the susceptibility measured by elastoresistivity scales to the renormalized nematic susceptibility obtained from shear modulus’ softening. This observation challenges present theories of resistivity anisotropy and electronic nematicity in iron pnictides.


Our results extend the current picture of the family of iron-based superconductors revealing that LaFeAsO as representative for the ’1111’-type compounds shows a softening of the shear modulus similar to what has been observed in '122'-~\cite{fernandes2010,bending122} and '11'-~\cite{bending11} compounds. Thereby, adding our results for the ’1111’-system make the observed softening of the shear modulus an ubiquitous property of iron-based superconductors. Stark qualitative differences of the behaviour in LaFeAsO to the observations in the other pnictide families however pose new questions on the elastoresistivity, calling for further theoretical Investigation. 


\begin{acknowledgements}
We thank C. Meingast and A. Böhmer for valuable discussions. Support by Deutsche Forschungsgemeinschaft (DFG) under Germany’s Excellence Strategy EXC2181/1-390900948 (the Heidelberg STRUCTURES Excellence Cluster) as well as through the Priority Programme SPP1458 via grant no. KL1827/6-1 is gratefully acknowledged. L.W. acknowledges funding through grants no. WA4313/1-1\&2.
\end{acknowledgements}

\hfill

\appendix
\section*{Appendix}\label{Sec_Appendix}

\renewcommand{\thefigure}{A\arabic{figure}}
\setcounter{figure}{0} 

\begin{figure}[h!]
	\includegraphics[width=1.0\columnwidth,clip] {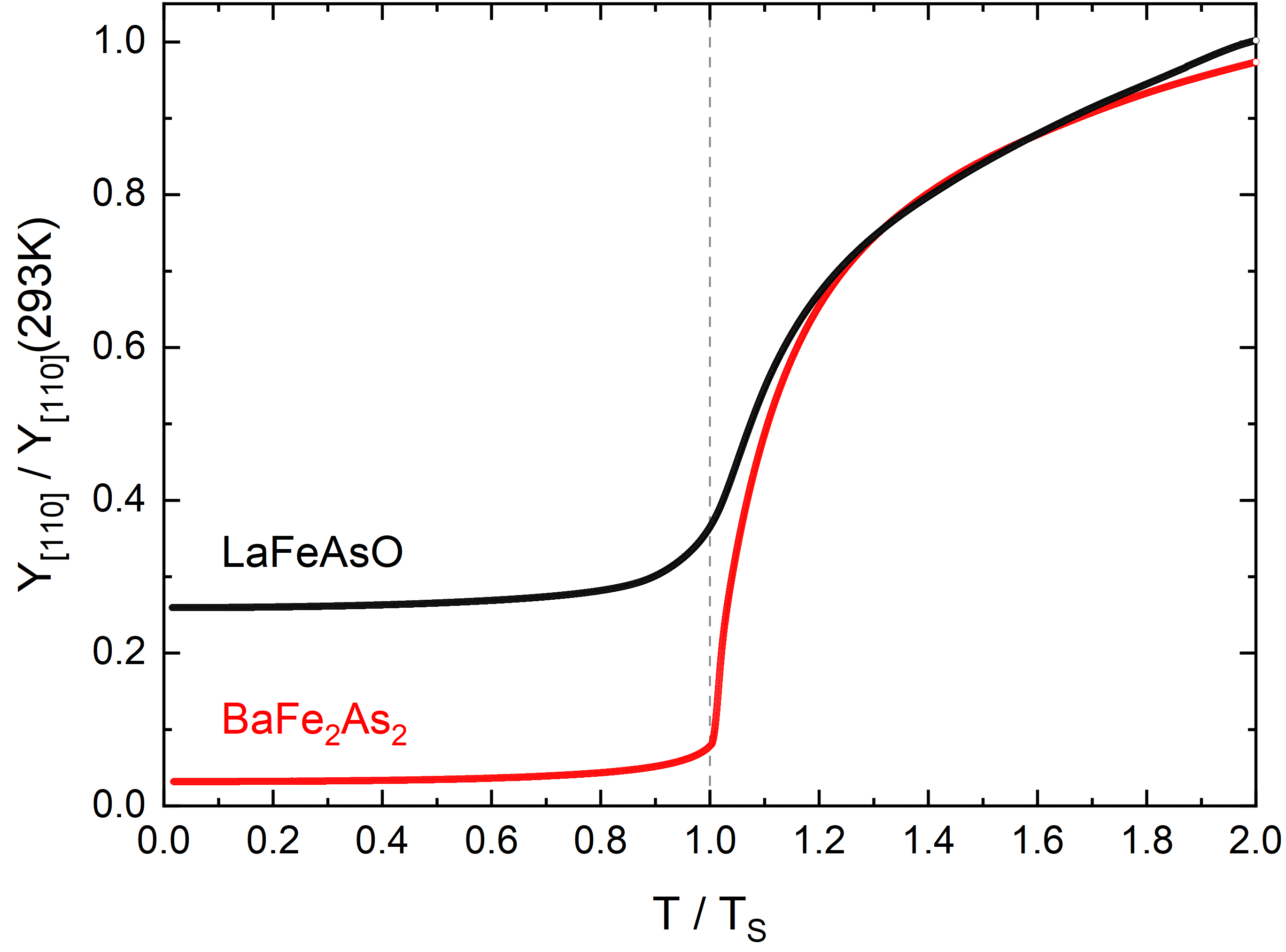}
	\caption{Normalized Young's modulus in \la\ and \ba\ $vs.$ temperature scaled to the respective structural transition temperature \ts.}\label{Y_scaled}
\end{figure}

\begin{figure}[h!]
	\includegraphics[width=1.0\columnwidth,clip] {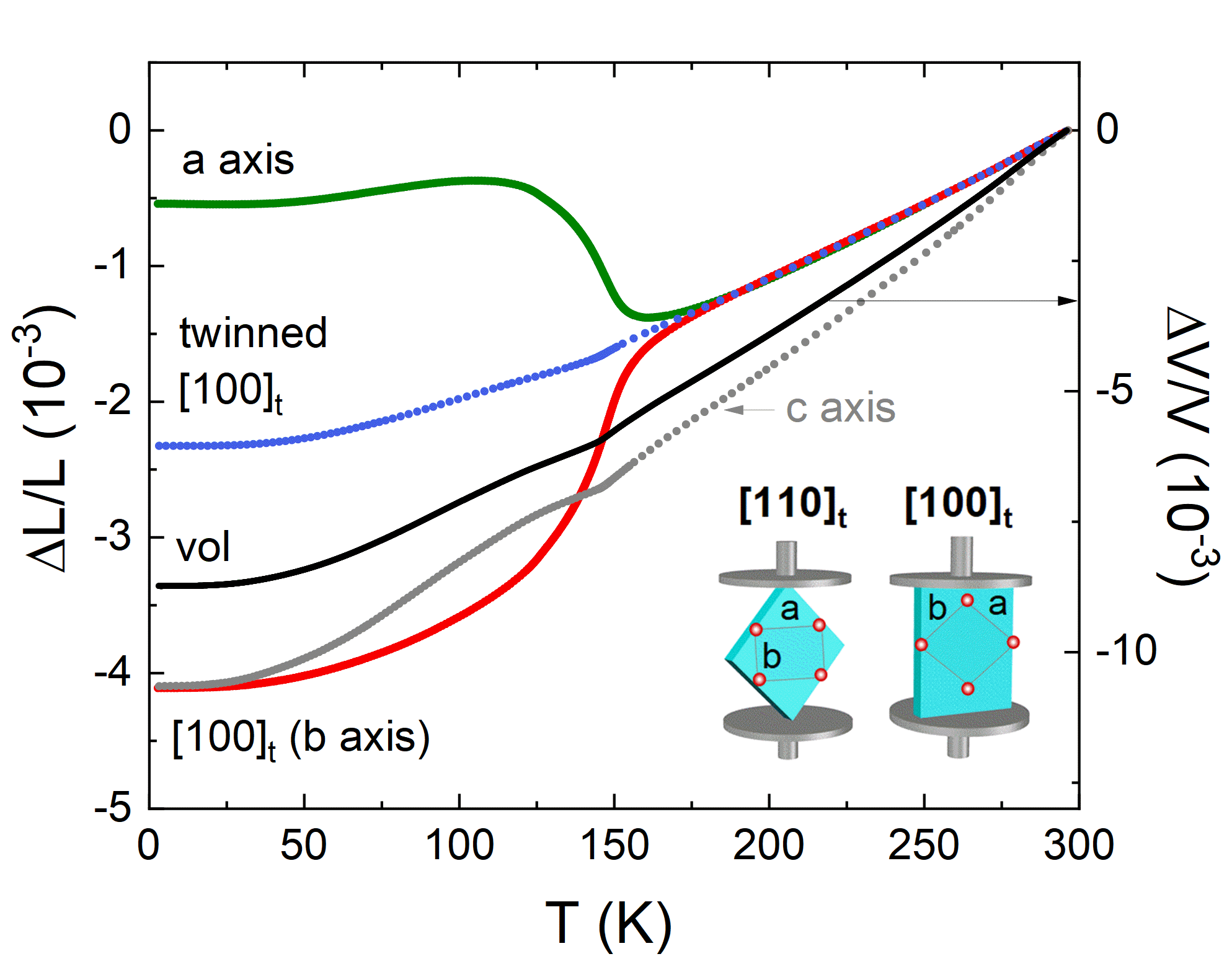}
	\caption{\rev{Relative length changes $\Delta L/L$ of the three main crystallographic axes and as measured along the twinned [100]$_{t}$ direction as well as the resultant calculated volume changes $\Delta V/V$. The inset schematically illustrates the measurements along the twinned and detwinned in-plane directions from which the $a$ axis length changes are obtained.} 
}\label{twinned}
\end{figure}

\hfill \newpage

\bibliography{nematicity}

\end{document}